\documentclass[10pt]{article}
\usepackage[dvips]{graphicx}
\usepackage{tabularx}
\usepackage{amsmath,amssymb}
\usepackage{bm}
\newcommand{\eq}{\begin{equation}}
\newcommand{\eqend}{\end{equation}}
\newcommand{\ovl}{\overline}
\newcommand{\A}{\alpha}
\newcommand{\B}{\beta}

 \addtocounter{section}{0}
 \makeatletter \@addtoreset{equation}{section}
 \makeatother

\title{Supersymmetric Yang-Mills Theory on the Noncommutative Geometry}
\author{Satoshi Ishihara, 
  \footnote{E-mail:satoshi@yukawa.kyoto-u.ac.jp}
\and
Hironobu Kataoka,
  \footnote{E-mail:s499756@hyogo-c.ed.jp}
\and
Atsuko Matsukawa,
  \footnote{E-mail:Atsuko Matsukawa@cap.ocn.ne.jp}
\and 
Hikaru Sato,
  \footnote{E-mail:hikaru\underline{ }sato@gakushikai.jp} \\
{\it Department of Physics, Hyogo University of Education} \\
{\it Shimokume, Kato-shi, Hyogo 673-1494, Japan} \\
\\ 
Masafumi Shimojo \footnote{E-mail:shimo0@ei.fukui-nct.ac.jp} \\
{\it Department of Electronics and Information Engineering, }\\
{\it Fukui National College of Technology,} \\
{\it Geshicho, Sabae-Shi, Fukui 916-8507, Japan}
}
\begin{document}
\maketitle
\begin{abstract}
Recently, we found the supersymmetric counterpart of the spectral triple. 
When we restrict the representation space to the fermionic functions of matter fields, 
the counterpart which we name "the triple" reduces to the original spectral triple which defines noncommutative 
geometry. We see that the fluctuation to the supersymmetric Dirac operator induced by algebra in the 
triple generates vector supermultiplet which mediates gauge interaction. 
Following the supersymmetric version of 
spectral action principle, 
we calculate the heat kernel expansion of the square of fluctuated Dirac operator and obtain the 
correct supersymmetric Yang-Mills action with U(N) gauge symmetry.
\end{abstract}
\section{INTRODUCTION}
\ \ \ \ 
The standard model of high energy physics coupled to gravity was derived on the basis of 
noncommutative  geometry(NCG) by Connes and his co-workers\cite{connes1,connes2,connes4}. 
Their result was that if the space-time was a 
product of a continuous Riemann manifold $M$ and a finite space $F$ of KO-dimension 6, gauge theories of the 
standard model were uniquely derived\cite{connes3}.  

The framework of a NCG is specified by a set called spectral triple\cite{connes0}. Let it be denoted by   
$(\mathcal{H}_0,\mathcal{A}_0,\mathcal{D}_0)$. Here, $\mathcal{A}_0$ is a noncommutative complex algebra, acting 
on the Hilbert space $\mathcal{H}_0$, whose elements correspond to spinorial wave functions of physical matter fields, 
while the Dirac operator $\mathcal{D}_0$ is a self adjoint operator with compact resolvent. The operator 
plays the role of the inverse of infinitesimal unit of length $ds$ of ordinary geometry and 
satisfies the condition that $[\mathcal{D}_0,a]$ is bounded $\forall a \in \mathcal{A}_0$.
$Z/2$ grading $\gamma$ and the real structure $\mathcal{J}$ were taken into account to determine the KO dimension. These axioms are given in the 
Euclidean signature\cite{connes7}.

The automorphisms of the algebra $\mathcal{A}_0$ are separated into equivalence classes under 
its normal subgroup. In the same way, the space of metrics, i.e. Dirac operators has a foliation of equivalence classes and the internal fluctuation of a metric is given as follows:
\begin{equation}
\tilde{\mathcal{D}}_0 = \mathcal{D}_0 + A+ JAJ^{-1},\ A=\sum a_i[\mathcal{D},b_i],\ a_i,b_i\in \mathcal{A}.
\end{equation}   
For the Dirac operator in the manifold $\mathcal{D}_{0M}=i\gamma^\mu\nabla_\mu\otimes 1$, the fluctuation 
$A+JAJ^{-1}$ gives the gauge vector field, while for the Dirac operator in the finite space,$\mathcal{D}_{0F}$, it 
gives the Higgs field\cite{connes7,connes10}. 

The action of the NCG model is obtained by the spectral action principle and expressed by
\begin{equation}
\langle \psi \tilde{\mathcal{D}}_0  \psi \rangle+ {\rm Tr}(f(P)), \label{totalaction0}
\end{equation}
where $\psi$ is a fermionic field which is an element 
in $\mathcal{H}_0$, $f(x)$ is an auxiliary smooth function on a compact Riemannian manifold 
without boundary of dimension 4. The second term of the action (\ref{totalaction0}) represents the bosonic part 
obtained by the principle, which asserts that it depends only on the spectrum 
of the squared Dirac operator $P=\tilde{\mathcal{D}}_0^2$\cite{connes8}. It includes not only 
non-abelian gauge theory but also theory of Higgs field and Einstein's general relativity.

In our last paper\cite{paper0}, we extended the spectral triple defined on the flat Riemannian manifold to a counterpart in the 
supersymmetric theory which may overcome various shortcomings of the standard model\cite{martin}, for example, 
hierarchy problem, many free parameters to be determined by experiments, lacking of unification of  running gauge 
coupling constants of renormalization group. 
We referred to the supersymmetric counterpart as symply "the triple" and the triple on the manifold was denoted by 
$(\mathcal{A}_M,\mathcal{H}_M,\mathcal{D}_M)$. 
Faithfully obeying the original idea that the Hilbert space of the spectral triple constituted of 
wave functions of matter fields while gauge fields and Higgs fields were derived from fluctuations 
of the Dirac operator, a component of the functional space $\mathcal{H}_M$ in our counterpart was made up 
by spinor and scalar functions of $C^\infty(M)$ which constructed a chiral or an antichiral supermultiplet 
and represented wave functions of a matter field and its superpartners. A Z/2 grading of the space $\mathcal{H}_M$ was 
given by the chirality of the supersymmetry transformation. So, the functional space in the manifold, $\mathcal{H}_M$, 
was a direct sum of $\mathcal{H}_+$ of chiral supermultiplets  and $\mathcal{H}_-$ of antichiral supermultiplets.  
The algebra $\mathcal{A}_M$ was also separated to two subsets $\mathcal{A}_+$ and $\mathcal{A}_-$ 
each of which was represented on $\mathcal{H}_+$ and $\mathcal{H}_-$, respectively. 
On the other hand, the representation of the Dirac operator $\mathcal{D}_M$ was defined on the whole $\mathcal{H}_M$. 
While the ingredients of NCG  was constructed in the Euclidean signature, 
the above construction was performed in the Minkowskian signature in order to incorporate supersymmetry.
We note that the triple did not define a new NCG. However, 
projecting $\mathcal{H}_M$ to the fermionic part $\mathcal{H}_{0M}$ and changing the signature 
from Minkowskian to the Euclidean one by the Wick rotation, we found that it reduced to the theory 
constructed on the original spectral triple.

In addition to the discussion of the triple defined on the manifold, we should also intorduce the counterpart  
in the finite space denoted by $(\mathcal{A}_F,\mathcal{H}_F,\mathcal{D}_F )$ in order to incorporate gauge 
quantum numbers of matter particles and their superpartners and to obtain mass terms of them. 

In this paper, we will calculate the spectral action using the introduced Dirac operators following the 
supersymetric version of spectral action principle.  In section 2, we review the construction of the supersymmetric 
counterpart extended from the original spectral triple on the manifold. We also introduce the counterpart on the 
finite space so that the triple defined on $M\otimes F$ will be given. 
In section 3, we calculate the internal fluctuation of the Dirac operator which induces the vector 
supemultiplet with $U(N)$ internal degrees of freedom. 
We can define an adequate supersymmetric invariant product of elements in $\mathcal{H}_M$ and   
obtain bilinear form similar to the first term in (\ref{totalaction0}). It represents the action 
for the chiral and antichiral supermultiplets of matter fields and their superpartners.  
We will calculate the square of the fluctuated Dirac operator and Seelay-DeWitt coefficients of heat kernel expansion 
so that we will arrive at the correct action of the supersymmetric Yang-Mills theory.
\section{\large SUPERSYMMETRICALLY EXTENDED TRIPLE}
\ \ \ 
The supersymmetric counterpart of the spactral triple on the flat Riemannian manifold $M$ has been introduced in our last paper. 
In this chapter, let us review it and at the same time introduce 
the counterpart on the finite space $F$ so that the triple 
$(\mathcal{A},\mathcal{H},\mathcal{D})$ extended from the spectral triple defined on $M\otimes F$ can be discussed.
The functional space $\mathcal{H}$ is the product denoted by
\begin{equation}
\mathcal{H} =\mathcal{H}_M\otimes \mathcal{H}_F. \label{H}
\end{equation}
The functional space on the the Minkowskian space-time manifold 
is the direct sum of two subsets, $\mathcal{H}_+$ and 
$\mathcal{H}_-$:
\begin{equation}
\mathcal{H}_M = \mathcal{H}_+ \oplus \mathcal{H}_- \label{HM}
\end{equation}
The element of $\mathcal{H}_M$ is given by
\begin{align}
\Psi  & = \begin{pmatrix}
\Psi_+ \\
\Psi_-
\end{pmatrix}
=\Phi_+ + \Phi_-, 
\label{Phi} \\
& \Phi_+=
\begin{pmatrix}
\Psi_+ \\
0^3 
\end{pmatrix} \in \mathcal{H}_+,\ 
\Phi_-= \begin{pmatrix}
0^3 \\
\Psi_-
\end{pmatrix}
\in \mathcal{H}_-.  
\end{align} 
Here, $\Psi_+$, $\Psi_-$ are denoted by 
\begin{equation}
(\Psi_+)_i = (\varphi_+(x),\psi_{+\alpha}(x),F_+(x))^T,\ i=1,2,3,  \label{Psi+}
\end{equation}
and 
\begin{equation}
(\Psi_-)_{\bar{i}} = (\varphi_-(x),\psi_-^{\dot{\alpha}}(x),F_-(x))^T,\ \bar{i}=1,2,3, \label{Psi-}
\end{equation}
in the vector notation. Here, $\varphi_+$ and $F_+$ of $\Psi_+$ are complex scalar functions with mass dimension 1 and 
2,respectively, and $\psi_\alpha$,$\alpha=1,2$ are the Weyl spinors on the space-time $M$ which 
have mass dimension $\frac{3}{2}$  and transform as the $(\frac{1}{2},0)$ represenation of 
the Lorentz group, $SL(2,C)$. 
$\Psi(x)$ obey the following chiral supersymmetry transformation and 
form a chiral supermultiplet.
\begin{equation}
\begin{cases}
\delta_\xi\varphi_+ =\sqrt{2}\xi^\alpha \psi_{+\alpha}, \\
\delta_\xi\psi_{+\alpha} = i\sqrt{2}\sigma^\mu_{\alpha\dot{\alpha}}\bar{\xi}^{\dot{\alpha}}\partial_\mu\varphi_+
+\sqrt{2}\xi_\alpha F_+,\\
\delta_\xi F_+ = i\sqrt{2}\bar{\xi}_{\dot{\alpha}}\bar{\sigma}^{\mu\dot{\alpha}\alpha}\partial_\mu\psi_{+\alpha}
\end{cases}. \label{deltaxi}
\end{equation}

On the other hand, $\psi^{\dot{\A}}$ transform as 
the $(0,\frac{1}{2})$ of $SL(2,C)$ and $\Psi_-(x)$ form a antichiral supermultiplet which obey the 
antichiral supersymmetry transformation as follows:
\begin{equation}
\begin{cases}
\delta_\xi\varphi_- = \sqrt{2}\bar{\xi}_{\dot{\alpha}}\psi^{\dot{\alpha}}_-, \\
\delta_\xi\psi_-^{\dot{\alpha}} = i\sqrt{2}\bar{\sigma}^{\mu\dot{\alpha}\alpha}\xi_\alpha\partial_\mu\varphi_-
+\sqrt{2}\bar{\xi}^{\dot{\alpha}}F_-, \\
\delta_\xi F_- = i\sqrt{2}\xi^\alpha\sigma^\mu_{\alpha\dot{\alpha}}\partial_\mu\psi_-^{\dot{\alpha}}. 
\end{cases} \label{deltabxi}
\end{equation}

A Z/2 grading of the functional space $\mathcal{H}_M$ is given by an operator $\gamma_M$ in $\mathcal{H}_M$ which is 
defined by
\begin{equation}
\gamma_M =\begin{pmatrix}
-i & 0\\
0 & i
\end{pmatrix}, 
\end{equation}
on the basis such that $\gamma_M(\Psi_+)=-i$ and $\gamma_M(\Psi_-)=i$. 

Let us discuss the space $\mathcal{H}_F$. 
$\mathcal{H}_F$ is the space with the basis of the labels 
$q_L^a$ and $q_R^a$, which correspond to some matter particles and their 
superpartners, such as quarks, squarks and auxiliary fields.  
Here $a$ is the index, $a=1,\dots N$, in order to 
introduce internal degrees of freedom. $L$ and $R$ denote the eigenstates of 
the $Z/2$ grading $\gamma_F$, which is defined by
\begin{equation}
\gamma_F = \begin{pmatrix}
-1 & 0\\
0 & 1
\end{pmatrix},
\end{equation}
in the basis of $\mathcal{H}_F$ given by
\begin{equation}
Q^a =\begin{pmatrix}
q_L^a \\
q_R^a
\end{pmatrix} \in\mathcal{H}_F,
\end{equation} 
In this basis, we have $\gamma_F(q_L^a)=-1$, and $\gamma_F(q_R^a)=1$. 
The wave functions of the supermaltiplets in $\mathcal{H}$ are expressed by 
$(\Psi_+,\Psi_-)\otimes (q_L^a,q_R^a)$. In order to evade fermion 
doubling \cite{Gracia,lizzi}, we impose that the physical 
wave functions obey the following condition:
\begin{equation}
\gamma =\gamma_M\gamma_F =i.
\end{equation}
Then for the supermultiplet which is a set of a left-handed 
fermionic matter field and its superpartner and auxiliary field, 
we have 
\begin{align}
\Psi_L^a & = q_L^q\otimes \Phi_+ \nonumber \\
& = q_L^a \otimes (\varphi_+,\psi_{+\alpha},F_+,0^3)^T, \label{PsiL}
\end{align} 
in the Minkowskian signature and the physical wave functions of the supermultiplet amount to
\begin{align}
q_{L\alpha}^a & = q_L^a\otimes \psi_{+\alpha}(x), \\
\tilde{q}_L^a(x) & = q_L^a\otimes \varphi_+(x), \\
F_L^a(x) & = q_L^a\otimes F_+(x). 
\end{align} 
For the wave functions of the  right-handed matter field, we have
\begin{equation}
\Psi_R^a(x) = q_R^a\otimes \Phi_- = q_R^a\otimes (0^3, \varphi_-^\ast,\bar{\psi}_-^{\dot{\alpha}},F_-^\ast)^T
\end{equation}
and
\begin{align}
q_R^{a\dot{\alpha}}(x) & = q_R^a \otimes \bar{\psi}_-^{\dot{\alpha}}(x),\\
\tilde{q}_R^a (x) & = q_R^a \otimes \varphi_-(x)^\ast, \\
F_R^a(x) & = q_R^a(x) \otimes F_-(x)^\ast.  \label{FR}
\end{align}

For the state $\Psi \in \mathcal{H}_M$ , the charge conjugate state $\Psi^c$ is given by
\begin{equation}
\Psi^c = \begin{pmatrix}
\Psi_+^c\\
\Psi_-^c
\end{pmatrix}.
\end{equation}
The antilinear operator $\mathcal{J}_M$ is defined by
\begin{equation}
\Psi^c = \mathcal{J}_M\Psi =C\Psi^*,
\end{equation}
so that it is given by
\begin{equation}
\mathcal{J}_M = C\otimes *,
\end{equation}
where C is the following charge conjugation matrix:
\begin{equation}
C =\left(\begin{array}{ccc|ccc}
 & & & 1 & &  \\
 & \mathbf{0} & & & \epsilon_{\alpha\beta} & \\
   & & & & & 1  \\
  \hline 
  1 & & & & &  \\
  & \epsilon^{\dot{\alpha}\dot{\beta}} & & & \mathbf{0} & \\
   & & 1 & & &
\end{array}\right),
\end{equation}
and $*$ is the complex counjugation. The operator $\mathcal{J}_M $ obeys the 
following relation:
\begin{equation} 
\mathcal{J}_M\gamma_M =\gamma_M\mathcal{J}_M.
\end{equation}
The real structure $J_M$ is now expressed for the basis 
of the Hilbert space $(\Phi,\Phi^c)^T$ in the following form:
\begin{equation}
J_M = \begin{pmatrix}
0 & \mathcal{J}_M^{-1}\\
\mathcal{J}_M & 0
\end{pmatrix}. 
\end{equation}
The Z/2 grading $\Gamma_M$ on the basis is expressed by
\begin{equation}
\Gamma_M = \begin{pmatrix}
\gamma_M & 0\\
0 & \gamma_M
\end{pmatrix}
\end{equation}

In the finite space, the antilinear operator $\mathcal{J}_F$ is defined by
\begin{equation}
\mathcal{J}_F = \begin{pmatrix}
0 & 1\\
1 & 0
\end{pmatrix} \otimes * .
\end{equation} 
Then the anti-matter particle supermultiplet $Q_a^c$ is related to $Q_a$ by
\begin{equation}
Q_a^c = \mathcal{J}_F Q^a.
\end{equation}
On the basis $(Q_a,Q_a^c)^T$, the real structure $J_F$ and the Z/2 grading 
is expressed as follows:
\begin{align}
J_F =\begin{pmatrix}
0 & \mathcal{J}_F^{-1}\\
\mathcal{J}_F & 0
\end{pmatrix},\ 
\Gamma_F =
\begin{pmatrix}
\gamma_F & 0\\
0 & \gamma_F
\end{pmatrix}.
\end{align} 
   
Corresponding to the construction of the functional space (\ref{H}),(\ref{HM}), 
the algebra $\mathcal{A}$ represented by them are expressed as
\begin{align}
\mathcal{A} & = \mathcal{A}_M\otimes \mathcal{A}_F, \\
\mathcal{A}_M & = \mathcal{A}_+ \oplus \mathcal{A}_-.
\end{align}
Here an element of $\mathcal{A_+}$, $u_a$, which acts on $\mathcal{H}_+$, and an element of 
$\mathcal{A}_-$,$\bar{u}_a$, which acts on $\mathcal{H}_-$ are given by 
\begin{align}
(u_a)_{ij} & = \frac{1}{m_0}
\begin{pmatrix}
\varphi_a & 0 & 0\\
\psi_{a\alpha} & \varphi_a & 0 \\
F_a & -\psi_a^\alpha & \varphi_a
\end{pmatrix} \in \mathcal{A}_+, 
\label{ua}
\\
(\bar{u}_a)_{\bar{i}\bar{j}} & = \frac{1}{m_0}
\begin{pmatrix}
\varphi_a^\ast & 0 & 0\\
\bar{\psi}_a^{\dot{\alpha}} & \varphi_a^\ast & 0 \\
 F_a^\ast & -\bar{\psi}_{a\dot{\alpha}} & \varphi_a^\ast
\end{pmatrix} \in \mathcal{A}_-, \label{barua},
\end{align}
where $\{\varphi_a(\varphi_a^\ast), \psi_{a\alpha}(\bar{\psi}_a^{\dot{\alpha}}),F_a(F_a^\ast)\}$
are chiral(antichiral) multiplet but are not necessarily elements in (\ref{Psi+})(\ref{Psi-}).

The total supersymmetric Dirac operator $D$ is defined as follows:
\begin{equation}
D = D_M\otimes 1 + \Gamma_M\otimes D_F, \label{DAtot}
\end{equation}
On the basis $(\Phi,\Phi^c)^T$, the operator $D_M$ is given by 
\begin{equation}
D_M = \begin{pmatrix}
\mathcal{D}_M & 0\\
0 & \mathcal{J}_M \mathcal{D}_M \mathcal{J}_M^{-1}
\end{pmatrix},
\end{equation}
and
\begin{equation}
\mathcal{D}_M = -i\begin{pmatrix}
0 & \bar{\mathcal{D}}_{i\bar{j}} \\
\mathcal{D}_{\bar{i}j} & 0
\end{pmatrix}, 
\label{DM0}
\end{equation}
where
\begin{equation}
\mathcal{D}_{ij}=\begin{pmatrix}
0 & 0 & 1\\
0 & i\bar{\sigma}^\mu\partial_\mu & 0\\
\Box & 0 & 0
\end{pmatrix},\ 
\bar{\mathcal{D}}_{i\bar{j}}= \begin{pmatrix}
0 & 0 & 1\\
0 & i\sigma^\mu\partial_\mu & 0\\
\Box & 0 & 0
\end{pmatrix}.
\label{DM}
\end{equation}

The supersymmetric invariant product in $\mathcal{H}_M$ is defined by 
\begin{equation}
(\Phi^\prime,\Phi) =\int_M d^4 x \Phi^{\prime\dagger}\Gamma_0 \Phi, \label{ssproduct}
\end{equation}
where $\Gamma_0$ is given by
\begin{equation}
\Gamma_0 =\begin{pmatrix}
0 & \mathit{\Gamma}_0\\
\mathit{\Gamma}_0 & 0
\end{pmatrix}, \label{Gamma0}
\end{equation}
and \begin{equation}
\mathit{\Gamma}_0 =\begin{pmatrix}
0 & 0 & 1\\
0 & -1 & 0\\
1 & 0 & 0
\end{pmatrix}. \label{itGamma0}
\end{equation}
%

On the basis $(Q^a,Q_a^c)$, the Dirac operator on the finite space $D_F$ 
is defined as follows:
\begin{equation}
D_F = \begin{pmatrix}
\mathcal{D}_F & 0\\
0 & \mathcal{J}_F\mathcal{D}_F\mathcal{J}_F^{-1},
\end{pmatrix}
\end{equation} 
and 
\begin{equation}
\mathcal{D}_F =
\begin{pmatrix}
0 & m^\dagger\\
m & 0
\end{pmatrix}, \label{DF}
\end{equation}
where $m$ is the mass matrix with respect to the family index.
The above formalism was given in the framework of Minkowskian signature in order to 
incorporate supersymmetry. 
When we restrict the functional space $\mathcal{H}$ to its fermionic part $\mathcal{H}_0$ and 
the algebra and Dirac operator to those on $\mathcal{H}_0$ transferring to the Euclidean signature, 
we can go back to the original spectral triple which gives the framework of NCG.

The Wick rotation transforms the space-time variables in Minkowskian coordinates to  
Euclidean ones as follows:
\begin{equation}
x^0 \rightarrow -ix^0. 
\end{equation}  
The algebra of $SL(2,C)$ turns out to be the algebra of $SU(2)\otimes SU(2)$ under the 
rotation. 
The Weyl spinors which transform as $(\frac{1}{2},0)$, $(0,\frac{1}{2})$ of $SL(2,C)$ 
are to be replaced by $(\frac{1}{2},0)$ and $(0,\frac{1}{2})$ represenations of 
$SU(2)\otimes SU(2)$,respectively. The spinors which have appeared in $\mathcal{H}_M$, 
 and $\mathcal{A}_M$ are replaced as follows:
\begin{align}
\psi_{(+)\alpha} \rightarrow \rho_{(+)\alpha}, 
& \ \psi_{(+)}^\alpha \rightarrow \rho_{(+)}^{\alpha\ast}, \label{replacepsi+}\\
\bar{\psi}_{(-)}^{\dot{\alpha}}\rightarrow \omega_{(-)}^{\dot{\alpha}}, & \ 
\bar{\psi}_{(-)\dot{\alpha}} \rightarrow \omega_{(-)\dot{\alpha}}^\ast,\label{replacepsi-}
\end{align}
where spinors with indices $\alpha$ transform as $(\frac{1}{2},0)$ and those with indices $\dot{\alpha}$ 
transform as $(0,\frac{1}{2})$ of $SU(2)\otimes SU(2) $,respectively.  
The upper index is related to the complex conjugate of the lower 
index by $\rho^1=\rho_2^\ast$, $\rho^2=-\rho_1^\ast$, $\omega^{\dot{1}}=\omega_{\dot{2}}^\ast$, 
$\omega^{\dot{2}}=-\omega_{\dot{1}}^\ast $. 
The metric and Pauli matrices which have appeared  in the Dirac operator are to be replaced by 
\begin{align}
g^{\mu\nu} = (-1,1,1,1) & \rightarrow \eta^{\mu\nu}=(1,1,1,1) \label{replacemetric}\\
\sigma_\mu  \rightarrow \sigma^\mu_E = (\sigma^0,i\sigma^i), & \ \  
\bar{\sigma}^\mu \rightarrow \bar{\sigma}_E^\mu =(\sigma^0,-i\sigma^i).\label{replacesigma}
\end{align}

Embedding these expressions (\ref{replacepsi+})$\sim$ (\ref{replacesigma}), the triple is  
rewritten in the Euclidean signature.
%
%
The basis of $\mathcal{H}_M$ is denoted by the same form as (\ref{Psi+}),(\ref{Psi-}) 
but now $\Psi_+$ and $\Psi_-$  are given by 
\begin{align}
(\Psi_+)_i & =(\varphi_+,\rho_\alpha,F_+), \\
(\Psi_-)_{\bar{i}} & =(\varphi_-^\ast ,\omega^{\dot{\alpha}},F_-^\ast),
\end{align}

The elements of $\mathcal{A}_M$  which correspond to (\ref{ua}) and (\ref{barua}) is now 
given by 
\begin{align}
(u_a)_{ij} & = \frac{1}{m_0}
\begin{pmatrix}
\varphi_a & 0 & 0\\
\rho_{a\alpha} & \varphi_a & 0 \\
F_a & -\rho_a^\alpha & \varphi_a
\end{pmatrix} \in \mathcal{A}_+, 
\\
(\bar{u}_a)_{\bar{i}\bar{j}} & = \frac{1}{m_0}
\begin{pmatrix}
\varphi_a^\ast & 0 & 0\\
\omega_a^{\dot{\alpha}} & \varphi_a^\ast & 0 \\
F_a^\ast & -\omega_{a\dot{\alpha}} & \varphi_a^\ast
\end{pmatrix} \in \mathcal{A}_-.
\end{align}
%
For the Dirac operator on the Minkowskian manifold in 
(\ref{DM0}) and (\ref{DM}), we have
\begin{equation}
\mathcal{D}_M = -i\begin{pmatrix}
0 & \bar{\mathcal{D}}_E \\
\mathcal{D}_E & 0
\end{pmatrix}, \label{DME0}
\end{equation}
where
\begin{equation}
\mathcal{D}_{E \bar{i}j}=\begin{pmatrix}
0 & 0 & 1\\
0 & \bar{\sigma}_E^\mu\partial_\mu & 0\\
\Box_E & 0 & 0
\end{pmatrix},\ 
\bar{\mathcal{D}}_{E i\bar{j}}= \begin{pmatrix}
0 & 0 & 1\\
0 & \sigma_E^\mu\partial_\mu & 0\\
\Box_E & 0 & 0
\end{pmatrix}, \label{DME}
\end{equation}
with 
\begin{equation}
\Box_E = \eta^{\mu\nu}\partial_\mu\partial_\nu = \partial_0^2+\partial_i^2.
\end{equation}
The invariant product under Euclidean supersymmetry transformation 
is given by the 
same form as (\ref{ssproduct}), but $\Gamma_0$ should be replaced by
\begin{equation}
\Gamma_0 =\begin{pmatrix}
\mathit{\Gamma}_0 & 0\\
0 &\mathit{\Gamma}_0
\end{pmatrix} \label{EGamma}
\end{equation} 

\section{\large INTERNAL FLUCTUATION AND VECTOR SUPERMULTIPLET}
In the supersymmetric counterpart of the NCG, the vector superfield is to be introduced as the 
internal fluctuation to the Dirac operator $D$. 
\begin{equation}
D \rightarrow \tilde{D} = D + V + J V J^{-1},\ V=\sum_a U_a^\prime[D,U_a],\ U_a\in\mathcal{A},  
\end{equation}
where $J=J_M\otimes J_F$. 
We assume that $\mathcal{A}_F$ is the algebra of $N\times N$ complex matrix functions for the space of $Q^a$.
As the algebra $\mathcal{A}_M$ is a direct sum of $\mathcal{A}_+$ and $\mathcal{A}_-$,
we need two sets of elements,$\Pi_+$ and $\Pi_-$:
\begin{align}
\Pi_+ & = \{u_a:a=1,2,\cdots n\} \subset \mathcal{A}_+\otimes \mathcal{A}_F \\
\Pi_- & = \{\bar{u}_a:a=1,2,\cdots n\} \subset\mathcal{A}_-\otimes \mathcal{A}_F,
\end{align}
where $u_a$ and $\bar{u}_a$ are given in the matrix form of (\ref{ua}) and (\ref{barua}). 
In the space of $Q^a$,
they are also $N\times N$ complex matrix functions which act on internal degrees of freedom of $\mathcal{H}_F$

On the other hand, we assume that the algebra $\mathcal{A}$ for the space of antiparticles $Q_c^a$ is that of 
constant complex number $c$, so that the elements are unit matrix in $\mathcal{A}_M$. 
Then the fluctuation induced by the term $c[\mathcal{D}_M,c]$ vanishes. 
But the term $JVJ^{-1}$ carries the same non-vanishing fluctuation induced by $N\times N$ 
complex matrices in the space of $Q^a$ to the space of $Q_a^c$.       

Since the product of chiral(antichiral) supermultiplets is again the chiral(antichiral) supermultiplet, the 
elements of $\Pi_+$($\Pi_-$) are chosen such that products of two or more $u_a^\prime s(\bar{u}_as) $ 
do not belong to $\Pi_+$($\Pi_-$) any more.

We shall define the following scalar,spinor and vector superfields as the bilinear form of the 
two component functions in $u_a\in \Pi_+$ and $\bar{u}_a\in \Pi_- $;
\begin{align}
m_0^2 C & = \sum_a c_a\varphi_a^\ast \varphi_a, \label{m02C}\\
m_0^2 \chi_\alpha & = -i\sqrt{2}\sum_ac_a\varphi_a^\ast\psi_{a\alpha},\\
m_2^2(M+iN) & =-2i\sum_a c_a\varphi^\ast_aF_a,\\
m_0^2A_\mu & = -i\sum_a c_a[(\varphi_a^\ast\partial_\mu\varphi_a-\partial_\mu\varphi_a^\ast\varphi_a)
-i\bar{\psi}_{a\dot{\alpha}}\bar{\sigma}^{\dot{\alpha}\alpha}_\mu\psi_{a\alpha}], \\
m_0^2\lambda_\alpha & = \sqrt{2}i\sum_a c_a(F_a^\ast\psi_{a\alpha}
-i\sigma^\mu_{\alpha\dot{\alpha}}\bar{\psi}_a^{\dot{\alpha}}\partial_\mu\varphi_a), \\
m_0^2 D & = \sum_a c_a[2F_a^\ast F_a -2(\partial^\mu\varphi_a^\ast\partial_\mu\varphi_a) \nonumber \\
& +i\{\partial_\mu\bar{\psi}_{a\dot{\alpha}}\bar{\sigma}^{\mu\dot{\alpha}\alpha}\psi_{a\alpha}
-\bar{\psi}_{a\dot{\alpha}}\bar{\sigma}^{\mu\dot{\alpha}\alpha}\partial_\mu\psi_{a\alpha}\}], \label{m02D}
\end{align}
where $c_a$ are the real coefficients. Using (\ref{deltaxi}) and (\ref{deltabxi}), 
we can show that these fields have the transformation property of the vector 
supermultiplet expressed by
\begin{align}
\delta_\xi C & = i\xi^\alpha \chi_\A -i\bar{\xi}_{\dot{\A}}\bar{\chi}^{\dot{\A}}, \\
\delta_\xi \chi_\alpha & = -i\sigma^\mu_{\A\dot{\A}}\bar{\xi}^{\dot{\A}}(-A_\mu+i\partial_\mu C)+\xi_\A(M+iN),\\
\frac{1}{2}\delta_\xi(M+iN) & = \bar{\xi}_{\dot{\A}}(\bar{\lambda}^{\dot{\A}}+i\bar{\sigma}^{\mu\dot{\A}\A}\partial_\mu \chi_\A),\\
\delta_\xi A^\mu & = i\xi^\A\sigma^\mu_{\A\dot{\A}}\bar{\lambda}^{\dot{\A}}
+ i\bar{\xi}_{\dot{\A}}\bar{\sigma}^{\mu\dot{\A}\A}\lambda_\A+\xi^\A\partial^\mu \chi_\A
+\bar{\xi}_{\dot{\A}}\partial_\mu\bar{\chi}^{\dot{\A}}, \\  
\delta_\xi \lambda_\A & = \sigma_\alpha^{\mu\nu\B}\xi_\B(\partial_\mu A_\nu-\partial_\nu A_\mu)+i\xi_\A D,\\
\delta_\xi D & = -\xi^\A\sigma^\mu_{\A\dot{\A}}\partial_\mu\bar{\lambda}^{\dot{\A}}
+\bar{\xi}_{\dot{\A}}\bar{\sigma}^{\mu\dot{\A}\A}\partial_\mu\lambda_\A.
\end{align}
If we express these fields as the superfield, we have
\begin{align}
V(x,\theta,\bar{\theta}) & = C +\theta^\alpha(i\chi_\alpha)
+\bar{\theta}_{\dot{\alpha}}(-i\chi^{\dot{\alpha}}) 
+\theta^\alpha\sigma^\mu_{\alpha\dot{\alpha}}\bar{\theta}^{\dot{\alpha}}(-A_\mu) \nonumber \\
& +\theta\theta\left[\frac{i}{2}(M+iN)\right]+\bar{\theta}\bar{\theta}\left[-\frac{i}{2}(M-iN)\right] \nonumber \\
& +\theta\theta\bar{\theta}_{\dot{\alpha}}\left[i(\bar{\lambda}^{\dot{\alpha}}
+\frac{i}{2}\bar{\sigma}^{\mu\dot{\alpha}\alpha}\partial_\mu\chi_\alpha)\right]
+\bar{\theta}\bar{\theta}\theta^\alpha\left[-i(\lambda_\alpha
+\frac{i}{2}\sigma^\mu_{\alpha\dot{\alpha}}\partial_\mu\bar{\chi}^{\dot{\alpha}})\right] \nonumber\\
& + \theta\theta\bar{\theta}\bar{\theta}(\frac{1}{2}D+\frac{1}{4}\Box C).
\end{align} 

When we define the vector superfield(\ref{m02C})$\sim$ (\ref{m02D}), there is an ambiguity 
due to the choice of the algebraic elements. In order to see this, we consider two arbitrary 
elements of the algebra given by $u_0\in \mathcal{A}_+$ and $\bar{u}_0 \in \mathcal{A}_-$. It 
turns out that the following functions obtained by these elements obey the supersymmetry 
transformation of the vector supermultiplet.
\begin{align}
m_0 C_0 & = \varphi_0+\varphi_0^\ast, \\
m_0\chi_{0\alpha} & =-i\sqrt{2}\psi_{0\alpha}, \\
m_0(M_0+iN_0) & = -2iF_0,\\
m_0A_{0\mu} & = -i\partial_\mu(\varphi_0-\varphi_0^\ast),\\
\lambda_{0\alpha} & =0,\\
D_0 & =0.
\end{align}
Then we can redefine $C,\chi_\alpha,M,N$ such that
\begin{align}
C  & \rightarrow C+C_0 = 0, \\
\chi_\alpha & \rightarrow \chi_\alpha+\chi_{0\alpha} =0, \\
M+iN & \rightarrow (M+M_0)+i(N+N_0)=0. 
\end{align}
To choose $C$,$\chi_\alpha$, $M$ and $N$ in the vector supermultiplet to be zero is called the 
Wess-Zumino gauge. This gauge is realized in (\ref{m02C})$\sim$ (\ref{m02D}) by the 
following condition:
\begin{align}
\sum_a c_a\varphi_a^\ast \varphi_a & = 0, \nonumber \\
\sum_a c_a\varphi_a^\ast \psi_a^\alpha & =0, \label{WessZumino}\\
\sum_a c_a\varphi_a^\ast F_a & =0.\nonumber 
\end{align}
Hereafter let us call (\ref{WessZumino}) the Wezz-Zumino condition.

Since $u_a$ and $\bar{u}_a$ are $N\times N$ complex matrix functions,
$A_\mu$,$D$, $\lambda_\alpha$ are also $N\times N$ complex matrix functions and 
parametrized by
\begin{align}
A_\mu(x) & = \sum_{l=0}^{N^2-1} A_\mu^l(x)\frac{T_l}{2}, \label{Amu} \\
D(x) & = \sum_{l=0}^{N^2-1} D^l(x)\frac{T_l}{2}, \label{Dx}\\       
\lambda_\alpha(x) & = \sum_{l=0}^{N^2-1} \lambda_\alpha^l(x)\frac{T_l}{2}. \label{lambda}
\end{align}
Here, $T^l$ are basis of generators which belong to fundamental representation of the Lie algebra associated 
with Lie group $U(N)$ and normalized as follows:
\begin{equation}
{\rm Tr}(T_aT_b) =2\delta_{ab}.
\end{equation}  
Since $A^\mu(x)$ and $D(x)$ are hermitian so that $A_\mu^l(x)$ and $D^l(x)$ are real functions. On the other hand, 
$\lambda_\alpha^l(x)$ are complex functions.

The supersymmetric Dirac operator modified by the fluctuation is denoted by  
\begin{equation}
\tilde{\mathcal{D}}_M = -i\begin{pmatrix}
0 & \tilde{\bar{\mathcal{D}}}_{i\bar{j}} \\
\tilde{\mathcal{D}}_{\bar{i}j} & 0
\end{pmatrix}.
\label{tDM}
\end{equation}

We consider the fluctuation due to $u_a\in \Pi_+$ and $\bar{u}_a\in \Pi_-$. 
The contribution to $\tilde{\mathcal{D}}_{\bar{i}j}$ is given by 
the following form:
\begin{align}
V_{\bar{i}j} & = -2 \sum_a c_a (\bar{u}_a)_{\bar{i}\bar{k}}
[i\mathcal{D}_M,u_a]_{\bar{k}j} \nonumber \\
& = -2\sum_ac_a (\bar{u}_a)_{\bar{i}\bar{k}}\mathcal{D}_{\bar{k}l}(u_a)_{lj}
\end{align}
and the contribution to $\tilde{\bar{D}}_{i\bar{j}} $ is given by
\begin{align}
\bar{V}_{i\bar{j}} & = 2\sum_a c_a(u_a)_{ik}[i\mathcal{D}_M,\bar{u}_a]_{k\bar{j}} \nonumber \\
& = 2\sum_ac_a (u_a)_{ik}\bar{\mathcal{D}}_{k\bar{l}}(\bar{u}_a)_{\bar{l}\bar{j}}.
\end{align}
We shall calculate in the Wess-Zumino gauge. Using the definition of the 
vector supermultiplet given by (\ref{m02C})$\sim$ (\ref{m02D}), we obtain the following result:
\begin{equation}
V_{\bar{i}j} = -\begin{pmatrix}
0 & 0 & 0\\
i\sqrt{2}\bar{\lambda}^{\dot{\alpha}} & -\bar{\sigma}^{\mu\dot{\alpha}\alpha}A_\mu & 0\\
D+i\partial^\mu A_\mu+2iA_\mu\partial^\mu & i\sqrt{2}\lambda^\alpha & 0
\end{pmatrix},
\end{equation}
and
\begin{equation}
\bar{V}_{i\bar{j}}  = \begin{pmatrix}
0 & 0 & 0\\
-i\sqrt{2}\lambda_\alpha & \sigma^\mu_{\alpha\dot{\alpha}}A_\mu & 0\\
D-i\partial^\mu A_\mu-2iA_\mu\partial^\mu & -i\sqrt{2}\bar{\lambda}_{\dot{\alpha}} & 0
\end{pmatrix}.
\end{equation}

There are the additional fluctuation due to $u_{ab}=u_au_b\in\mathcal{A}_+$ and 
$\bar{u}_{ab} = \bar{u}_a\bar{u}_b\in \mathcal{A}_- $, where $a,b=1,\cdots,n$. This fluctuation 
is not contained in the fluctuation due to $u_a\in \Pi_+$ and $\bar{u}_a\in \Pi_- $ since 
$u_{ab}\notin\Pi_+ $ and $\bar{u}_{ab}\notin\Pi_- $. The component of fields of $u_{ab}$ are 
expressed by the matrix form of (\ref{ua}) and (\ref{barua}) and each field is given by
\begin{equation}
u_{ab} = \{\varphi_{ab},\psi_{ab\alpha},F_{ab}\},
\end{equation}
where 
\begin{align}
\varphi_{ab} & = \frac{1}{m_0}\varphi_a\varphi_b, \label{phiab}\\
\psi_{ab\alpha} & = \frac{1}{m_0}(\psi_{a\alpha}\varphi_b+\varphi_a\psi_{b\alpha}), \\
F_{ab} & =\frac{1}{m_0}(\varphi_a F_b+F_a\varphi_b-\psi_a^\alpha\psi_{b\alpha}). \label{fab}
\end{align}
The component fields of $\bar{u}_{ab}$ are the complex conjugate functions of 
(\ref{phiab})$\sim$(\ref{fab}).

It turns out that 
the gauge-covariant form of $\tilde{\mathcal{D}}_M$ is obtained by adding the following 
fluctuation due to $u_{ab}$ and $\bar{u}_{ab}$:
\begin{align}
V_{\bar{i}j}^\prime & = 2 \sum_{a,b} c_ac_b (\bar{u}_{ab})_{\bar{i}\bar{k}}
[i\mathcal{D}_M,u_{ab}]_{\bar{k}j} \nonumber \\
& = 2\sum_{a,b}c_ac_b (\bar{u}_{ab})_{\bar{i}\bar{k}}\mathcal{D}_{\bar{k}l}(u_{ab})_{lj}
\end{align}
and 
\begin{align}
\bar{V}_{i\bar{j}}^\prime & = 
2\sum_{a,b} c_ac_b(u_{ab})_{ik}[i\mathcal{D}_M,\bar{u}_{ab}]_{k\bar{j}} \nonumber \\
& = 2\sum_{a,b}c_ac_b (u_{ab})_{ik}\bar{\mathcal{D}}_{k\bar{l}}(\bar{u}_{ab})_{\bar{l}\bar{j}}.
\end{align}
Taking into account the Wess-Zumino gauge condition given by (\ref{WessZumino}), we obtain
\begin{equation}
V_{\bar{3}1}^\prime = -\bar{V}^\prime_{3\bar{1}} = -A_\mu A^\mu,
\end{equation}
and other matrix element turns out to be zero. The fluctuation 
due to higher order products of $u_a$ or $\bar{u}_a$ such as 
$u_{abc}=u_au_bu_c$ or $\bar{u}_{abc}= \bar{u}_a\bar{u}_b\bar{u}_c$ vanishes due to the 
Wess-Zumino gauge condition. Thus the total fluctuation in the Wess-Zumino gauge amounts to
\begin{align}
V_{\bar{i}j}^{WZ} & = V_{\bar{i}j}+V_{\bar{i}j}^\prime,\\
\bar{V}_{i\bar{j}}^{WZ} & = \bar{V}_{i\bar{j}}+\bar{V}_{i\bar{j}}^\prime,
\end{align}
and the Dirac operator with fluctuation denoted by (\ref{tDM})
is finally given by 
\begin{align}
\tilde{\mathcal{D}}_{\bar{i}j} & = \mathcal{D}_{\bar{i}j}+V_{\bar{i}j}^{WZ} \nonumber \\
& = \begin{pmatrix}
0 & 0 & 1\\
-i\sqrt{2}\bar{\lambda}^{\dot{\alpha}} & i\bar{\sigma}^\mu\mathcal{D}_\mu & 0\\
\mathcal{D}_\mu\mathcal{D}^\mu -D & -i\sqrt{2}\lambda^\alpha & 0
\end{pmatrix}, \label{Dij}
\end{align}
and 
\begin{align}
\tilde{\bar{\mathcal{D}}}_{i\bar{j}} & = \bar{\mathcal{D}}_{i\bar{j}}+\bar{V}_{i\bar{j}}^{WZ} \nonumber \\
& = \begin{pmatrix}
0 & 0 & 1\\
-i\sqrt{2}\lambda_\alpha & i\sigma^\mu\mathcal{D}_\mu & 0\\
\mathcal{D}_\mu\mathcal{D}^\mu +D & -i\sqrt{2}\bar{\lambda}_{\dot{\alpha}} & 0
\end{pmatrix}, \label{bDij}
\end{align}
where $\mathcal{D}_\mu$ is the covariant derivative,
\begin{equation}
\mathcal{D}_\mu = \partial_\mu -iA_\mu.
\end{equation}

As for the Dirac operator on the finite space, we assume that $\mathcal{D}_F$ in (\ref{DF}) has not the internal degrees 
of freedom so that the fluctuation for it does not arise.

The modified total Dirac operator on the basis $\Phi\otimes Q^a$ is given by
\begin{equation}
\tilde{\mathcal{D}}_{tot} = \tilde{\mathcal{D}}_M -i\gamma_M\otimes \mathcal{D}_F. \label{D0}
\end{equation} 
Let us see that  the counterpart of the first term in (\ref{totalaction0}), which is 
in our supersymmetric case the part of the spectral action for the matter particles and their 
superpartners, is given by the bilinear form of supersymmetric invariant product defined in (\ref{ssproduct}) 
with the total Dirac operator in (\ref{D0}).  
\begin{align}
I_{matter} & = \left(\Psi_L+\Psi_R,i\mathcal{D}_{tot}(\Psi_L+\Psi_R)\right) \nonumber \\
& = \left(\Psi_L+\Psi_R,i\mathcal{D}_M(\Psi_L+\Psi_R)\right)
+\left(\Psi_L+\Psi_R,\gamma_M\otimes \mathcal{D}_F (\Psi_L+\Psi_R)\right) \nonumber \\
& = (\Psi_L,\tilde{\mathcal{D}}\Psi_L)+(\Psi_R,\tilde{\ovl{\mathcal{D}}}\Psi_R)
+(\Psi_L,im^\dagger\Psi_R) -(\Psi_R,im\Psi_L), \label{Imatter}
\end{align}
where $\Psi_L$,$\Psi_R$ are left-handed and right-handed particle supermultiplets in 
(\ref{PsiL})$\sim$(\ref{FR}):
\begin{align}
\Psi_L^a & = (\tilde{q}_L^a,q_{L\alpha}^a,F_L^a,0^3),\\
\Psi_R^a & = (0^3,\tilde{q}_R^a,q_R^{a\dot{\alpha}},F_R^a).
\end{align}
Using the definition of supersymmetric invariant product (\ref{ssproduct}), (\ref{Gamma0}), 
(\ref{itGamma0}) and fluctuated Dirac operator (\ref{tDM}),(\ref{Dij}),(\ref{bDij}), 
the kinetic part of the matter particles are expressed by
\begin{align}
I_L & =  (\Psi_L,\tilde{\mathcal{D}}\Psi_L)\nonumber \\
 & =  
\int_M d^4x \left[
-\mathcal{D}_\mu \tilde{q}_L^\ast \mathcal{D}^\mu \tilde{q}_L 
-i\bar{q}_{L\dot{\alpha}}\bar{\sigma}^{\mu\dot{\alpha}\alpha}\mathcal{D}_\mu q_{L\alpha} \right. \nonumber \\
& \left. 
-\sqrt{2}ig(\tilde{q}_L^\ast\lambda^\alpha q_{L\alpha}-\bar{q}_{L\dot{\alpha}}\bar{\lambda}^{\dot{\alpha}}\tilde{q}_L)
-g\tilde{q}_L^\ast D\tilde{q}_L+F_L^\ast F_L
\right], 
\end{align}
and \begin{align}
I_R & =  (\Psi_R, \tilde{\bar{\mathcal{D}}}\Psi_R)\nonumber \\
 & =  
\int_M d^4x \left[
-\mathcal{D}_\mu \tilde{q}_R^\ast \mathcal{D}^\mu \tilde{q}_R 
-i\bar{q}_{R\alpha}\sigma^{\mu}_{\alpha\dot{\alpha}}\mathcal{D}_\mu q_R^{\dot{\alpha}} \right. \nonumber \\
& \left. 
-\sqrt{2}ig(\tilde{q}_R^\ast\bar{\lambda}_{\dot{\alpha}} q_R^{\dot{\alpha}}-\bar{q}_R^\alpha\lambda_\alpha\tilde{q}_R)
-g\tilde{q}_R^\ast D\tilde{q}_R+F_R^\ast F_R
\right],
\end{align}
where $g$ is a rescale factor to the vector superfields that we will introduce later. 

As for the mass terms,i.e,the last two terms in (\ref{Imatter}), we redefine the phase of $\Psi_L$ as 
$\Psi_L\rightarrow i\Psi_L$, then 
we have 
\begin{align}
I_{mass} & = (\Psi_R,m\Psi_L) + {\rm h.c.} \nonumber \\
& =\int_M d^4x
[\tilde{q}_R^\ast mF_L +F_R^\ast m\tilde{q}_L-\ovl{q}_R^\alpha m q_{L\alpha} +{\rm h.c.}].
\end{align}  
\section{\normalsize SPECTRAL ACTION PRINCIPLE AND SUPER YANG-MILLS ACTION}
\ \ \ \ 
Let us start the final task to derive the super Yang-Mills theory following the supersymmetric version of prescription  
for constructing NCG particle models. 
In our noncommutaitve geometric approach to supersymmetry, 
we show that the action for vector supermultiplet will be obtained by the coefficients of heat 
kernel expansion of elliptic operator $P$:
\begin{equation}
Tr_{L^2}f(P) \simeq \sum_{n\geq 0}c_na_n(P),\label{hke}
\end{equation}
where $f(x)$ is an auxiliary smooth function on a smooth compact Riemannian manifold without boundary of 
dimension 4 similar to the non-supersymmetric case.
Since the contribution to $P$ from the antiparticles is 
the same as that of the particles, we consider only the contribution from the particles. Then the 
elliptic operator $P$ in our case is given by the square of the Wick rotated Euclidean Dirac operator 
$\tilde{\mathcal{D}}_{tot}$ expressed by the same form as (\ref{D0}):
\begin{equation}
\tilde{\mathcal{D}}_{tot} = \tilde{\mathcal{D}}_M -i\gamma_M\otimes \mathcal{D}_F, \label{DE0}
\end{equation}
where $\tilde{\mathcal{D}}_M$ is obtained from (\ref{tDM}),(\ref{Dij}) and (\ref{bDij}) but with the replacement 
of (\ref{replacemetric}),(\ref{replacesigma}).
Note that as for the case in which internal fluctuation to $\mathcal{D}_F$ exists, 
we will discuss in our next paper\cite{ishihara}.

The elliptic operator $P$ is expanded into the following form:
\begin{equation}
P = -(\eta^{\mu\nu}\partial_\mu\partial_\nu +\mathbb{A}^\mu\partial_\mu +\mathbb{B}). \label{P}
\end{equation}
The heat kernel coefficients $a_n$ in (\ref{hke}) are found in \cite{gilkey}. 
They vanish for n odd, and the first three $a_n$'s for $n$ even in the flat space are given by
\begin{align}
a_0(P) & =\frac{1}{16\pi^2}\int_M d^4x {\rm tr}_V(\mathbb{I}), \label{a0} \\
a_2(P) & =\frac{1}{16\pi^2}\int_M d^4x {\rm tr}_V(\mathbb{E}), \label{a2} \\
a_4(P) & =\frac{1}{32\pi^2}\int_M d^4x {\rm tr}_V(\mathbb{E}^2 \label{a4}
+\frac{1}{3}\mathbb{E}_{;\mu}^\mu+\frac{1}{6}\Omega_{\mu\nu}\Omega^{\mu\nu}), 
\end{align}
where $\mathbb{E}$ and the bundle curvature $\Omega^{\mu\nu}$ in the flat space are defined as follows:
\begin{align}
\mathbb{E} & = \mathbb{B} -(\partial_\mu \omega^\mu+\omega_\mu\omega^\mu), \label{E}\\
\Omega^{\mu\nu} & = \partial^\mu\omega^\nu -\partial^\nu\omega^\mu+[\omega^\mu,\omega^\nu], \label{Omega}\\
\omega^\mu & =\frac{1}{2}\mathbb{A}^\mu. \label{omega}
\end{align}

The coefficients $c_n$ in (\ref{hke}) depend on the functional form of $f(x)$. If $f(x)$ is flat near 0, 
it turns out that $c_{2k}=0$ for $k\geq 3$ and the heat kernel expansion terminates at $n=4$\cite{connes8}.

In (\ref{a0})$\sim$(\ref{a4}), ${\rm tr}_V$ denotes the trace over the vector bundle $V$. As for the supersymmetric theory 
we consider here, sections of the vector bundle $V$ are smooth functions bearing indices which correspond to 
internal and spin degrees of freedom of the chiral and antichiral supermultiplets. 
For the spin degrees of freedom, ${\rm tr}_V$ is the supertrace defined by
\begin{align}
{\rm Str} O & = \sum_i\langle i|(-1)^{2s}O|i\rangle \nonumber \\
   & = \sum_b\langle b|O|b\rangle -\sum_f \langle f|O|f\rangle,
\end{align}
where $s$ is the spin angular momentum and the states $|b\rangle$ and$|f\rangle$ stand for bosonic and fermionic state,
respectively. Being attached with spinor indices, a matrix $M$ represented on the space of supermultiplets span 
by the basis $(\varphi(x), \psi_\alpha(x), F(x))$ is expressed by  
\begin{equation}
M=
\begin{pmatrix}
M_{11} & M_{12}^\B & M_{13} \\
M_{21\A} & M_{22\A}^{\ \ \ \B} & M_{23\A} \\
M_{31} & M_{32}^\B & M_{33} 
\end{pmatrix}
\end{equation}
The supertrace of $M$ is given by
\begin{equation}
{\rm Str}M = {\rm tr}_V M_{11} -{\rm tr}_V M_{22\A}^{\ \ \ \A}+{\rm tr}_V M_{33}. \label{Str}
\end{equation}
The minus sign of tr$M_{22}$ is due to the supersymmetry. In order to consider the meaning of the sign, 
we rewrite the diagonal elements of $M$ on the superspace span by the basis $(\varphi, \theta^\B \psi_\B, \theta\theta F)$ 
as follows:
\begin{equation}
{\rm diag}\ M =(M_{11}, \theta^\alpha M_{22\A}^{\ \ \ \B}\frac{\partial}{\partial\theta^\B},M_{33}).
\end{equation} 
It is reasonable to assume that
\begin{equation}
{\rm tr}_V(\theta^\alpha M_{22\A}^{\ \ \ \B}\frac{\partial}{\partial\theta^\B}) =
-{\rm tr}_V(M_{22\A}^{\ \ \ \B}\frac{\partial}{\partial\theta^\B}\theta^\A) = -{\rm tr}_V M_{22\A}^{\ \ \ \A},
\end{equation}
so that (\ref{Str}) is established. In the same way, the supertrace of a matrix $\bar{M}$ represented on the space 
of antisupermultiplet $(\varphi^\ast,\bar{\psi}^{\dot{\alpha}},F^\ast)$ is given by
\begin{equation}
{\rm Str} \bar{M} = \bar{M}_{11} - \bar{M}_{{22}\dot{\alpha}}^{\ \dot{\alpha}}+\bar{M}_{33}.
\end{equation}

Now let us calculate the spectral action for $\tilde{\mathcal{D}}_{tot}^2$ where $\tilde{\mathcal{D}}_{tot}$ is 
given by (\ref{DE0}).
In the contribution to the spectral action from $\tilde{\mathcal{D}}_0^2$, the terms including $\mathcal{D}_F$ vanish 
since $\mathcal{D}_M$ anticommutes with $\gamma_M$ and 
\begin{equation}
{\rm Str}\gamma_M^2 ={\rm Str} 1 =0.
\end{equation} 
Thus, we consider the following elliptic operator $P$ in the Euclidean signature,
\begin{equation}
P = \tilde{\mathcal{D}}_M^2 =
\begin{pmatrix}
P_+ & 0\\
0 & P_-
\end{pmatrix}.
\end{equation}
By making use of the Wick-rotated expression of (\ref{Dij}) and (\ref{bDij}), $P_\pm$ amounts to
\begin{align}
P_+ & =-\tilde{\bar{\mathcal{D}}}_E\tilde{\mathcal{D}}_E \nonumber \\
& = -\begin{pmatrix}
\mathcal{D}_\mu\mathcal{D}^\mu -D & -i\sqrt{2}\lambda^\beta & 0\\
-i\sqrt{2}\sigma_{E\alpha\dot{\alpha}}^\mu\left((\mathcal{D}_\mu\bar{\lambda}^{\dot{\alpha}}) 
+\bar{\lambda}^{\dot{\alpha}}\mathcal{D}_\mu\right) & \mathcal{D}_\mu\mathcal{D}^\mu\delta_\alpha^\beta
+i\sigma_{E\alpha}^{\mu\nu\beta}F_{\mu\nu} & -i\sqrt{2}\lambda_\alpha \\
-2\bar{\lambda}_{\dot{\alpha}}\bar{\lambda}^{\dot{\alpha}} & 
-i\sqrt{2}\bar{\lambda}_{\dot{\alpha}}\bar{\sigma}_E^{\mu\dot{\alpha}\beta}\mathcal{D}_\mu &\mathcal{D}_\mu\mathcal{D}^\mu+D
\end{pmatrix}, \label{P+}\\
P_- & =-\tilde{\mathcal{D}}_E\tilde{\bar{\mathcal{D}}}_E \nonumber \\
& = -\begin{pmatrix}
\mathcal{D}_\mu\mathcal{D}^\mu +D & -i\sqrt{2}\bar{\lambda}_{\dot{\beta}} & 0\\
-i\sqrt{2}\bar{\sigma}_E^{\mu\dot{\alpha}\alpha}
\left((\mathcal{D}_\mu\lambda_\alpha) +\lambda_\alpha\mathcal{D}_\mu\right) 
& \mathcal{D}_\mu\mathcal{D}^\mu\delta_{\dot{\alpha}}^{\dot{\beta}}
+i\bar{\sigma}_{E\ \ \dot{\beta}}^{\mu\nu\dot{\alpha}}F_{\mu\nu} 
& -i\sqrt{2}\bar{\lambda}^{\dot{\alpha}} \\
-2\lambda^\alpha\lambda_\alpha & 
-i\sqrt{2}\lambda^\alpha\sigma^\mu_{E\alpha\dot{\beta}}\mathcal{D}_\mu &\mathcal{D}_\mu\mathcal{D}^\mu-D
\end{pmatrix}, \label{P-}
\end{align}
In (\ref{P+}) and (\ref{P-}), $\sigma_E^{\mu\nu}$ and $\bar{\sigma}_E^{\mu\nu}$ are defined by
\begin{align}
\sigma_E^{\mu\nu} & = (i\sigma^{0j},-\sigma^{ij}), \\
\bar{\sigma}_E^{\mu\nu} & = (i\bar{\sigma}^{0j},-\bar{\sigma}^{ij}),
\end{align} 
and 
\begin{align}
\sigma_\alpha^{\mu\nu\beta} & = \frac{1}{4}(\sigma^\mu_{\alpha\dot{\alpha}}\bar{\sigma}^{\nu\dot{\alpha}\beta}
-\sigma^\nu_{\alpha\dot{\alpha}}\bar{\sigma}^{\mu\dot{\alpha}\beta}), \\
\bar{\sigma}^{\mu\nu\dot{\alpha}}_{\ \ \ \dot{\beta}} & =\frac{1}{4}(
\bar{\sigma}^{\mu\dot{\alpha}\beta}\sigma^\nu_{\beta\dot{\beta}}
-\bar{\sigma}^{\nu\dot{\alpha}\beta}\sigma^\mu_{\beta\dot{\beta}}
) .
\end{align}

The field $A_\mu$, $\lambda_\alpha$, $D$ are the $N\times N$ matrices as shown in (\ref{Amu}),(\ref{Dx}),
(\ref{lambda}). They turn out to be the gauge, gaugino and auxiliary fields. The covariant derivative on 
spinors, say, $\lambda_\alpha$ is given by
\begin{equation}
\mathcal{D}_\mu = \partial_\mu \lambda_\alpha -i[A_\mu,\lambda_\alpha],
\end{equation}
and $F_{\mu\nu}$ is the field strength defined by
\begin{align}
F_{\mu\nu} & = i[\mathcal{D}_\mu,\mathcal{D}_\nu] \nonumber\\
          & = \partial_\mu A_\nu-\partial_\nu A_\mu -i[A_\mu,A_\nu].
\end{align}

We expand $P_{\pm}$ in the form given by (\ref{P}). Using the formulae given by (\ref{E}) and (\ref{omega}),
we obtain the following expressions:
\begin{align}
\mathbb{E}_+ & = \mathbb{B}_+ -(\partial_\mu\omega_+^\mu+\omega_{+\mu}\omega_+^\mu) =
\begin{pmatrix}
-D & -i\sqrt{2}\lambda^\beta & 0\\
\frac{-i}{\sqrt{2}}\sigma^\mu_{E\alpha\dot{\alpha}}(\mathcal{D}_\mu\bar{\lambda}^{\dot{\alpha}}) &
i\sigma_{E\alpha}^{\mu\nu\beta}F_{\mu\nu} & -i\sqrt{2}\lambda_\alpha \\
-2\bar{\lambda}_{\dot{\alpha}}\bar{\lambda}^{\dot{\alpha}} & 
\frac{i}{\sqrt{2}}(\mathcal{D}_\mu\bar{\lambda}_{\dot{\alpha}})\bar{\sigma}_E^{\mu\dot{\alpha}\beta} & D
\end{pmatrix} \label{E+} \\
\mathbb{E}_- & = \mathbb{B}_- -(\partial_\mu\omega_-^\mu+\omega_{-\mu}\omega_-^\mu) =
\begin{pmatrix}
D & -i\sqrt{2}\bar{\lambda}_{\dot{\beta}} & 0\\
\frac{-i}{\sqrt{2}}\bar{\sigma}_E^{\mu\dot{\alpha}\alpha}(\mathcal{D}_\mu\lambda_\alpha) &
i\bar{\sigma}_{E\ \ \dot{\beta}}^{\mu\nu\dot{\alpha}}F_{\mu\nu} & -i\sqrt{2}\bar{\lambda}^{\dot{\alpha}} \\
-2\lambda^\alpha\lambda_\alpha & 
\frac{i}{\sqrt{2}}(\mathcal{D}_\mu\lambda^\alpha)\sigma_{E\alpha\dot{\beta}}^\mu & -D
\end{pmatrix}. \label{E-}
\end{align}
The bundle curvature $\Omega_\pm^{\mu\nu}$ given by (\ref{Omega}) amounts to
\begin{align}
\Omega_+^{\mu\nu} & =\begin{pmatrix}
-iF^{\mu\nu} & 0 & 0\\
-\frac{i}{\sqrt{2}}[\sigma^\nu_{E\alpha\dot{\alpha}}
 (\mathcal{D}^\mu\bar{\lambda}^{\dot{\alpha}})
-\sigma^\mu_{E\alpha\dot{\alpha}}(\mathcal{D}^\nu\bar{\lambda}^{\dot{\alpha}})] & -iF^{\mu\nu}\delta_\alpha^\beta & 0\\
 0 & -\frac{i}{\sqrt{2}}
 [(\mathcal{D}^\mu\bar{\lambda}_{\dot{\alpha}})\bar{\sigma}_E^{\nu\dot{\alpha}\beta}-(\mathcal{D}^\nu \bar{\lambda}_{\dot{\alpha}})
 \bar{\sigma}_E^{\mu\dot{\alpha}\beta}] & -iF^{\mu\nu}
\end{pmatrix} \\
\Omega_-^{\mu\nu} & =\begin{pmatrix}
-iF^{\mu\nu} & 0 & 0\\
-\frac{i}{\sqrt{2}}[\bar{\sigma}^{\nu\dot{\alpha}\alpha}_{E}
 (\mathcal{D}^\mu\lambda_\alpha)
-\bar{\sigma}^{\mu\dot{\alpha}\alpha}_{E}(\mathcal{D}^\nu\lambda_\alpha)] & 
-iF^{\mu\nu}\delta_{\dot{\alpha}}^{\dot{\beta}} & 0\\
 0 & -\frac{i}{\sqrt{2}}
 [(\mathcal{D}^\mu\lambda^\alpha)\sigma_{E\alpha\dot{\beta}}^{\nu}-(\mathcal{D}^\nu \lambda^{\alpha})
 \sigma_{E\alpha\dot{\beta}}^{\mu}] & -iF^{\mu\nu}
\end{pmatrix}. 
\end{align}
From (\ref{E+}) and (\ref{E-}) we have
\begin{align}
{\rm Str}\mathbb{E}_+ & = {\rm Tr}[-D] -[-\sigma_{E\alpha}^{\mu\nu\alpha}]i{\rm Tr}F_{\mu\nu} +{\rm Tr}D =0
\label{strE1+},\\
{\rm Str}\mathbb{E}_- & =0, \label{strE1-}
\end{align}
since 
\begin{align}
\sigma_\alpha^{\mu\nu\alpha} & =
\frac{1}{4}(\sigma^\mu_{\alpha\dot{\alpha}}\bar{\sigma}^{\nu\dot{\alpha}\alpha} -
\sigma^\nu_{\alpha\dot{\alpha}}\bar{\sigma}^{\nu\dot{\alpha}\alpha}) 
=-\frac{1}{2}(g^{\mu\nu}-g^{\mu\nu}) =0, \\ 
\bar{\sigma}^{\mu\nu\dot{\alpha}}_{\ \ \ \dot{\alpha}} & =0.
\end{align}
As for the square of $\mathbb{E}^2$, we have
\begin{align}
{\rm Str} \mathbb{E}_+^2 & ={\rm Tr}[D^2-\lambda^\beta\sigma^\mu_{E\beta\dot{\beta}}
(\mathcal{D}_\mu\bar{\lambda}^{\dot{\beta}})] \nonumber \\
& - {\rm Tr}[\sigma^\mu_{E\alpha\dot{\alpha}}(\mathcal{D}_\mu\bar{\lambda}^{\dot{\alpha}})\lambda^\alpha
-\sigma_{E\alpha}^{\mu\nu\beta}\sigma_{E\beta}^{\lambda\kappa\alpha}F_{\mu\nu}F_{\lambda\kappa}
+\lambda_\alpha(\mathcal{D}_\mu\bar{\lambda}_{\dot{\beta}})\bar{\sigma}_E^{\mu\dot{\beta}\alpha} ] 
\nonumber \\
 & + {\rm Tr}[-(\mathcal{D}_\mu\bar{\lambda}_{\dot{\alpha}})\bar{\sigma}_E^{\dot{\alpha}\beta}\lambda_\beta
 +D^2] \nonumber \\
 & = {\rm Tr}[2D^2-4\bar{\lambda}_{\dot{\beta}}\bar{\sigma}^{\mu\dot{\beta}\beta}_E(\mathcal{D}_\mu\lambda_\beta)
 -F_{\mu\nu}F^{\mu\nu}-\frac{1}{2}\varepsilon^{\mu\nu\lambda\kappa}F_{\mu\nu}F_{\lambda\kappa}],
\label{strE2+}
\end{align}
and 
\begin{equation}
{\rm Str}\mathbb{E}_-^2 =
{\rm Tr}[2D^2 -4\bar{\lambda}_{\dot{\beta}}\bar{\sigma}_E^{\mu\dot{\beta}\beta}(\mathcal{D}_\mu\lambda_\beta )
-F_{\mu\nu}F^{\mu\nu}+\frac{1}{2}\varepsilon^{\mu\nu\lambda\kappa}F_{\mu\nu}F_{\lambda\kappa}],
\label{strE2-}
\end{equation}
where Tr denotes the trace over $N\times N$ matrices of the internal degrees of freedom. 
(\ref{strE2+}) and (\ref{strE2-}) give the following expression:
\begin{align}
{\rm tr}_V(\mathbb{E}^2) & = {\rm Str}\mathbb{E}_+^2 +{\rm Str}\mathbb{E}_-^2 \\
 & = 2{\rm Tr}[2D^2 -4\bar{\lambda}_{\dot{\beta}}\bar{\sigma}_E^{\mu\dot{\beta}\beta}(\mathcal{D}_\mu\lambda_\beta) -F_{\mu\nu}F^{\mu\nu}]
. \label{trVE2}
\end{align}
The supertrace of $\Omega_{\pm\mu\nu}\Omega_\pm^{\mu\nu}$ amounts to
\begin{equation}
{\rm Str}\Omega_{\pm\mu\nu}\Omega_\pm^{\mu\nu} ={\rm Tr}[-F_{\mu\nu}F^{\mu\nu}]-
{\rm Tr}[-F_{\mu\nu}F^{\mu\nu}{\bf 1}_2]+{\rm Tr}[-F_{\mu\nu}F^{\mu\nu}] =0.
\label{strOmega}
\end{equation}
%

Let us calculate the heat kernel coefficients. From (\ref{a0}), we obtain
\begin{equation}
a_0 =0, \label{cosmological}
\end{equation}
since the number of freedom of the bosonic sector is equal to the number of freedom of the 
fermionic sector due to the supersymmetry, so that Str$\mathbb{I}=0$. (\ref{cosmological}) indicates 
that the cosmological constant in the supersymmetric theory vanishes. 
From (\ref{strE1+}) and (\ref{strE1-}), the coefficients $a_2$ also vanishes,
\begin{equation}
a_2 =0.
\end{equation}
Finally, (\ref{trVE2}) and (\ref{strOmega}) give
\begin{equation}
a_4 =\frac{1}{16\pi^2} \int_Mdx^2
{\rm Tr}\left[
2D^2 -4\bar{\lambda}_{\dot{\beta}}\bar{\sigma}_E^{\mu\dot{\beta}\beta}(\mathcal{D}_\mu\lambda_\beta) -F_{\mu\nu}F^{\mu\nu}
\right],
\end{equation}
since ${\rm tr}_V(\mathbb{E}_{;\mu}^\mu) =0$. 

The Euclidean super Yang-Mills action $I_E$ is now given by
\begin{equation}
I_E = Tr_{L^2} f(\tilde{\mathcal{D}}_M^2) =f_4a_4.
\end{equation}
In order to obtain the physical action we change the signature back to 
the Minkowskian $\eta^{\mu\nu}\rightarrow g^{\mu\nu}$ with $\sigma_E\rightarrow i\sigma^\mu$ and 
rescale the vector supermultiplet as 
$\{A_\mu,\lambda_\alpha,D \}\rightarrow \{gA_\mu,g\lambda_\alpha,gD\}$, where $g$ turns out to be 
the gauge coupling constant. After this procedure we have the following super Yang-Mills action:
\begin{equation}
I_{SYM} = \int_Mdx^2
{\rm Tr}\left[
 -\frac{1}{2}F_{\mu\nu}F^{\mu\nu}- 2i\bar{\lambda}_{\dot{\beta}}\bar{\sigma}^{\mu\dot{\beta}\beta}(\mathcal{D}_\mu\lambda_\beta)  + D^2
\right],
\end{equation}
where we fixed the constant $f_4 $ such that 
\begin{equation}
\frac{f_4}{8\pi^2} =\frac{1}{g^2}.
\end{equation}
At last, we have arrived at the goal.

\section{\large CONCLUSIONS}
\ \ \ \ 
In this paper, we introduced the supersymmetric counterpart of the spectral triple which defines NCG 
on the finite space as well as on the manifold. 
Then the total Dirac operator was given in (\ref{DAtot}).
A vector supermultiplet was introduced as the internal fluctuation to the supersymmetrically extended 
Dirac operator $\mathcal{D}_M$ defined on the manifold. 
The modified Dirac operator $\tilde{\mathcal{D}}_M$ due to the fluctuation turns out to be 
supersymmetric and gauge-covariant if we introduce the internal degrees of freedom as the finite geometry. 
We considered the algebra of $N\times N$ complex matrices in the  finite space so that 
the internal fluctuation induce the vector supermultiplet with $U(N)$ gauge degrees of freedom. 

Following the prescription of NCG, we calculated the spectral action using our generalized supersymmetric 
Dirac operator $\tilde{\mathcal{D}}_{tot}$ given in (\ref{DE0}). 
The parts of the action which include kinetic terms and mass terms of matter fields 
and their superpartners are obtained from the supersymmetric invariant bilinear form. 
We calculated the coefficients of heat kernel expansion of the squared Dirac operator $P=\tilde{\mathcal{D}}_{tot}^2$. 
The terms including $\mathcal{D}_F$ did not contribute to them. 
We have found that the expansion coefficients for $\mathcal{D}_M^2$ successfully derived the super-Yang Mills action.  
As a result, we expressed the whole supersymmetric action for the matter supermultiplets and vector supermultiplets 
by the simple formula of the spectral action principle.

The method proposed in this paper to calculate the spectral action is applicable to 
the supersymmetric standard model. In this model, the Higgs bosons and their superpartners 
are introduced as the fluctuation to the supersymmetric Dirac operator on the finite space. 
The detailed discussions and calculations on this subject will be given in a separate paper\cite{ishihara}.  

\end{document}